# Understanding peculiarities in the optoelectronic characteristics of light emitting diodes based on (In,Ga)N/GaN nanowires

M. Musolino,[a]) A. Tahraoui, F. Limbach, J. Lähnemann, U. Jahn, O. Brandt, L. Geelhaar, and H. Riechert

*Paul-Drude-Institut für Festkörperelektronik, Hausvogteiplatz 5–7, D-10117 Berlin, Germany*

We investigate the effect of the p-type top contact on the optoelectronic characteristics of light emitting diodes (LEDs) based on (In,Ga)N/GaN nanowire (NW) ensembles grown by molecular beam epitaxy on Si substrates. We compare devices fabricated with either Ni/Au or indium tin oxide (ITO) top contact. The NW-LEDs with ITO exhibit a number density of NWs emitting electroluminescence about ten times higher, significantly lower turn-on voltage and series resistance, and a relative external quantum efficiency more than one order of magnitude higher than the sample with Ni/Au. These results show that limitations in the performance of such devices reported so far can be overcome by improving the p-type top-contact.

III-N nanowires (NWs) are an attractive alternative to conventional planar layers as the basis for light-emitting diodes (LEDs)[1–3] because they offer several conceptual advantages. The NW geometry enables the elastic relaxation of the strain induced by lattice mismatch at the free sidewalls,[4] thus permitting the growth of high quality (In,Ga)N/GaN heterostructures with high In content on Si substrates. Furthermore, the high aspect ratio of NWs inhibits the vertical propagation of extended defects,[5] and light extraction from arrays of NWs can be enhanced compared to planar devices.[2] In combination, these benefits could lead to cost-effective phosphorless monolithic white LEDs.[6]

In practice, LEDs based on GaN NW ensembles on Si substrates have been fabricated by several groups,[1,7–14] and significant limitations in device performance have been reported. In particular, careful investigations showed that only about 1 % of the NWs in the ensemble may emit electroluminescence (EL).[8,13,15] Also, in many cases high turn-on voltages in the range of 4.5–8 V were measured,[12,13,15,16] while for more complex NW structures lower values were obtained.[12,14,17] Thus, it seems fair to say that the actual implementation of the above conceptual advantages in device performance still remains to be demonstrated. Naturally, the processing of such LEDs is rather complex because of the three-dimensional morphology of NW ensembles. Therefore, it is at present unclear whether the reported limitations are peculiar to LEDs based on NW ensembles on Si substrates or such devices simply need further advances in processing technology.

One peculiarity of such NW-LEDs is that for typical device sizes they contain millions of NWs. Hence, the macroscopic LED actually consists of very many individual NW-LEDs contacted in parallel, and the overall device characteristics are determined by the properties of all the individual NW-LEDs. For example, NW-to-NW fluctuations in series resistance inevitably lead to a filamentation of the current path in the NW ensemble, and this phenomenon was in fact identified as the reason for the very low fraction of electroluminescent NWs.[13] Such fluctuations in series resistance could be caused either by non-uniformities of the top contact to the NW ensemble, by differences in the NW diameters, or by variations among the interfaces to the Si substrate. While the first two aspects can likely be improved, the last one may present an intrinsic limitation because it is associated with the fact that $Si_xN_y$ forms at the interface between GaN NWs and Si.[18,19]

In this letter, we assess whether the reported limitations in the device performance of LEDs based on (In,Ga)N/GaN NW ensembles on Si substrates are intrinsic in nature or can be overcome with improved processing. To this end, we compare devices processed from the same wafer with two different p-type top contacts. We conclusively demonstrate that both the low fraction of electroluminescent NWs and the high turn-on voltages are due to the properties of the p-type top contact, and thus do not indicate an intrinsic limitation.

The (In,Ga)N/GaN NW-LED structure employed in this work was grown by molecular beam epitaxy on an n-doped Si(111) substrate with the help of self-assembly processes. The n-type GaN base is doped with Si, while the p-type segment consists of a roughly 120-nm-thick GaN cap doped with Mg; a more detailed description of the layer stacking scheme can be found elsewhere.[13] The NW-LED samples were planarized by spin coating using a solution of hydrogen silsesquioxane (HSQ), which was subsequently transformed into solid $SiO_x$. Excess $SiO_x$ on the NW tips was removed by dry etching with $CHF_3$ until a considerable number of tips were uncovered. We aimed at removing about 70 nm of $SiO_x$ down from the tips of the highest NWs, but because of fluctuations in height of both NWs and $SiO_x$ layer, the actually etched thickness fluctuates on a local scale as well. Subsequently, two different top contacts were deposited: on one sample we sputtered 5 nm of Ni and consecutively 5 nm of Au (a common contact scheme[1,2,8,11,13]), while on the other one a 120-nm-thick ITO layer was sputtered directly onto the tips of the NWs. In order to decrease the resistivity of the ITO layer, it was annealed at 300 °C in $N_2$ atmosphere for 30 min. Optical characterization of two equivalent layers sputtered on glass showed that the 120-nm-thick ITO has a transmittance about two times higher than the Ni/Au layer. Finally, Ti/Au bonding pads and the Al/Au n-type contact were deposited on the top contacts and on the back side of the Si substrate, respectively.

---

[a])Author to whom correspondence should be addressed. Electronic mail: musolino@pdi-berlin.de

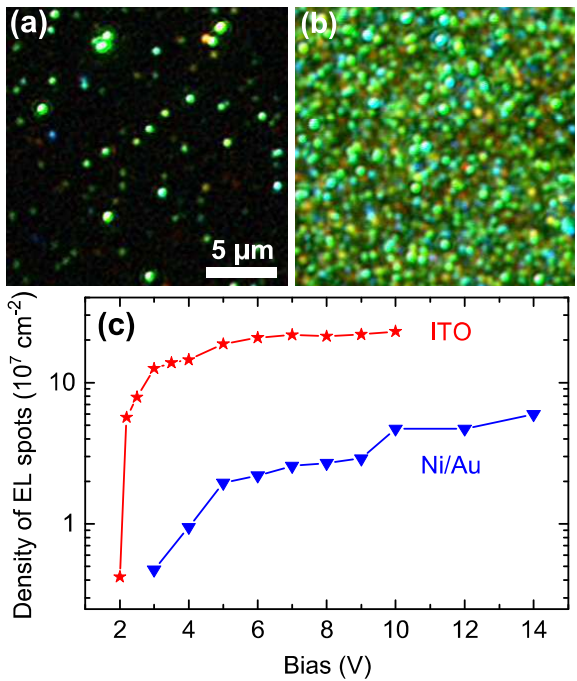

FIG. 1. (Color online) Top-view EL maps acquired at 6 V forward bias through an optical microscope for a NW-LED processed with (a) Ni/Au and (b) ITO. (c) Dependence of the number density of EL spots on the forward bias for these two different types of NW-LED.

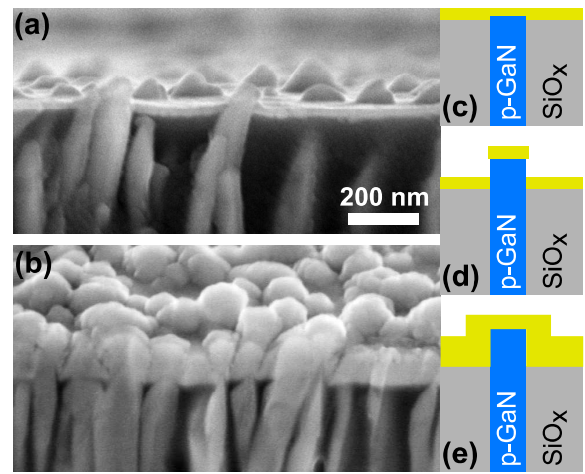

FIG. 2. (Color online) Bird's eye view SEM images of NW-LEDs processed with (a) Ni/Au and (b) ITO top contact, respectively. (c–e) Schematics of three possible configurations of the top contact.

As the first step to investigate the influence of the two types of top contact on device performance, we acquired electroluminescence maps of the samples with ITO and Ni/Au. Figures 1(a) and (b) show the resulting images acquired through an optical microscope on a device with an area of 0.2 mm$^2$ when a forward bias of 6 V was applied. Both samples exhibit a spotty emission pattern but the number density of EL spots is very different for the two samples. In particular, the sample with ITO exhibits a number density of emitting spots which is one order of magnitude higher than the sample with Ni/Au. In agreement with previous findings,[8,13,15] the EL map of the sample with Ni/Au contact is characterized by EL spots separated by large portions of surface area that are much darker, essentially without any emission of light. In contrast, the emitting NWs in the sample with ITO cover the whole surface without any dark areas in between. These results already prove that the low density of emitting NWs reported so far is due to a poor p-type top contact and not to limitations intrinsic to GaN NWs grown on Si, such as the formation of Si$_x$N$_y$ at the interface between the NWs and the substrate.

We acquired EL maps of the two samples for various forward biases; the extracted number densities of EL spots are summarized in Fig. 1(c). Several important differences between the two samples are remarkable. In particular, the turn-on of the first individual NW-LEDs occurs at lower forward bias in the device with ITO than in the one with Ni/Au. This result clearly proves that the ITO contact provides a lower series resistance. More interestingly, in the sample with ITO the number density of emitting NWs rises very quickly with bias, and almost saturates beyond 5 V. In contrast, in the sample contacted with Ni/Au the density of EL spots increases with the forward bias much more slowly than in the sample with ITO. Also compared to previous reports for similar type of NW-LEDs,[8,13,15] the highest density of EL spots counted at 10 V in the sample contacted with ITO is about one order of magnitude higher. We have to note that resolving the size of the EL spots is limited by the diffraction at the optical lens of the microscope to an area of approximately 0.2 μm$^2$. Therefore, roughly eight NWs might contribute to the formation of one single spot, and the density of spots is likely underestimated. Taking this effect into account, the actual density of NW-LEDs that emit EL could turn out to be of the same order of magnitude as the total density of NWs (which is about $5 \times 10^9$ cm$^{-2}$).

One might think that, because of fluctuations in the properties of the as-grown NW ensemble, a different number of individual NWs is actually capable of emitting light in the two pieces used for the fabrication of the devices. As discussed in the supplemental material, this is not the case.[20] It might also be possible that the Ni/Au layer contacts only few NWs, while the majority of them is not connected to the metal layer, and therefore cannot emit any electroluminescence. However, cross-sectional electron beam induced current (EBIC) measurements show that almost all the NWs are contacted.[13] Therefore, the causes of this drastic difference in number density of emitting spots must be sought in the different electrical properties of the two types of contacts. For instance, a lower contact resistance ($R_C$) between the tips of the individual NWs and the contacting layer and/or a more isotropic distribution of the resistance of the contacting layer ($R_L$) might be responsible for the observed effect.

In order to elucidate this question we consider the morphology of the two top contacts. Figure 2(a) and (b) show bird's eye view scanning electron microscopy (SEM) images of the NW-LED samples processed with Ni/Au and ITO con-tacts, respectively. Because of the different film thicknesses the contact area between the tips of the NWs and the contacting layers is very different in the two cases. As sketched in Fig. 2(c) and (d), the 10-nm-thick Ni/Au layer contacts either the top facet or a small portion of the side walls of the NW tips, depending on their height. If a long NW segment protrudes from the $SiO_x$ layer as indicated in Fig. 2(d), the amount of deposited Ni/Au is too small to form a continuous film on the sidewalls to the top facet. Moreover, the poor wetting behavior of Ni and Au on $SiO_x$ may produce voids and cracks in the metal film, thus reducing the contact area at the metal/NW interface.[21] In contrast, the transparent ITO layer can be made much thicker (120 nm in our case) without affecting the extraction of light significantly, and is hence able to fill any gaps between the NW tips so that the p-type top segments are completely embedded as depicted in Fig. 2(e), thus providing a larger contact area both with the top facet and the side walls of the NWs. The larger contact area can reduce $R_C$, which is inversely proportional to this parameter. To quantify the extent of this effect we calculated the contact area in these three possible cases. Using the typical dimension of our structures, we found that the contact area sketched in Fig. 2(e), $i.e.$ when the NW tips are completely surrounded by the contact material, is 4 and 12 times larger than in the cases depicted in Fig. 2(c) and (d), respectively. In addition, the larger thickness of the ITO layer can likely decrease $R_L$ and create a more uniform contact, allowing in this way the current to reach every part of the surface with equal ease.

Next, we present the analysis of the electrical characteristics of the two types of NW-LEDs. Figure 3 shows the current-voltage (I-V) curves of NW-LEDs with ITO and Ni/Au contact in linear scale. Five different devices, processed at the same time, were measured for each type of contact, and the presented data are typical ones. For the NW-LED with ITO, the current rises at much smaller biases than for the NW-LED with Ni/Au (see the supplemental material for a discussion of the leakage behavior and the graph in semi-logarithmic scale[20]). In particular, the former LED reaches the current of 5 mA for a forward bias of 3.3 V, while the latter LED requires 8.7 V to reach the same value. This value of the turn-on voltage found for the NW-LED contacted with ITO is lower than what was previously reported for similar NW-LEDs employing either Ni/Au/ITO[16] or Ni/ITO[15] as p-type contact. This result does not imply that the actual turn-on voltages of the individual NW-LEDs contacted with ITO are much smaller than the ones contacted with Ni/Au. In fact, as visible in the graph of Fig. 1(c), the first individual NW-LEDs are already emitting EL at 2 and 3 V in the devices with ITO and Ni/Au, respectively. What causes the large differences in the two I-V curves is mainly the different number of individual NW-LEDs in operation, and the actual amount of current that each NW conducts. As discussed before, in the devices contacted with ITO, the number of individual NW-LEDs in operation might be several tens of times greater than in the one with Ni/Au. At the same time, the lower $R_C$ produced by the larger contact area in the sample with ITO can significantly increase the current

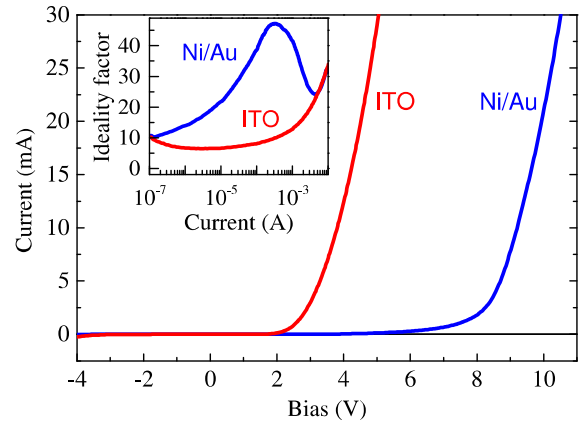

FIG. 3. (Color online) Current-voltage characteristics on a linear scale of NW-LEDs with a size of $0.2\,\text{mm}^2$ contacted with either Ni/Au or ITO. The inset shows the ideality factor derived from the I-V curves of the two samples versus the current.

through each individual NW-LED.

The series resistance ($R_S$) of the overall device was extracted from a linear fit to the I-V curve in the high bias range. The $R_S$ of the sample with ITO is approximately 30 % lower than that of the sample with Ni/Au: the absolute values are 30 and 43 $\Omega$, respectively. In addition, we calculated the ideality factor ($n$) of our NW-LEDs from the I-V curves using the formula $n = (q/kT)[(\delta \ln I/\delta V)^{-1}]$, where $q$, $k$, and $T$ are the elementary charge, the Boltzmann constant and the temperature, respectively. The results are depicted in the inset of Fig. 3. In the sample with Ni/Au, the ideality factor varies continuously over the entire range of currents, with values always higher than 10. In contrast, the sample with ITO exhibits an ideality factor approximately constant over the low-to-moderate current range ($10^{-6}$–$10^{-3}$ A), with an average value of about 9. This value is lower than the ones previously reported both for this kind of device and for the NW-LEDs obtained through top-down methods from planar GaN templates.[22,23] Values of $n$ much higher than 2 have also been measured for GaN-based planar LEDs, and as one possible origin poor quality of the p-type contact has been identified,[24] which could explain the observed difference in $n$ between the two types of devices considered here.

Last, we report the opto-electronic properties of the two samples. Figures 4(a) and (b) depict the room temperature EL spectra acquired for the samples with Ni/Au and ITO contacts at different current densities ($J$). The EL spectra of the two samples are composed of several peaks among which the main two are centered around 520 and 570 nm. The relative intensity of these two peaks is different in the samples with Ni/Au and ITO. This effect might be mainly explained by different current densities in the individual NWs emitting EL. In order to estimate the average current densities in the single NW-LEDs ($J_{NW}$) we used the formula $J_{NW} = J/(D_{on}A_{NW})$, where $A_{NW}$ and $D_{on}$ are the mean area of the NW top facets and the density of NWs emitting EL extrapolated from the number of EL spots counted before.

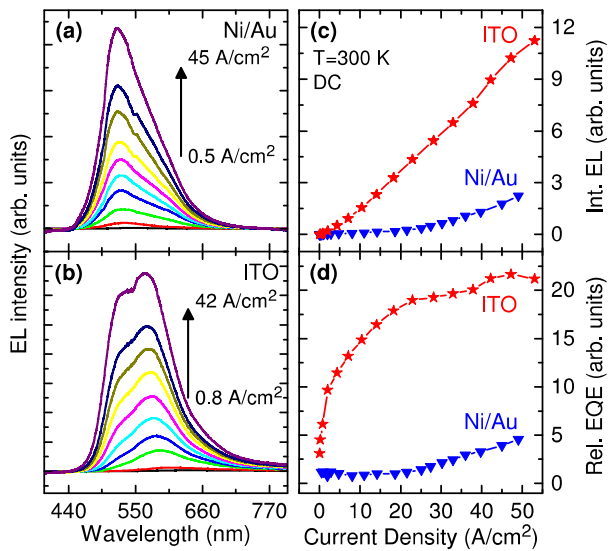

FIG. 4. (Color online) Room temperature EL spectra acquired at different current densities for the NW-LEDs with (a) Ni/Au and (b) ITO top-contact (same devices as in Fig. 3). (c) Integrated EL and (d) relative EQE plotted versus the injected direct current density.

We found that the ranges of total current densities considered in Fig. 4(a) and (b) roughly correspond to the following ranges of $J_{NW}$: 440–5600 A/cm$^2$ and 120–1300 A/cm$^2$ for the samples contacted with Ni/Au and ITO, respectively. If the limitation in optical resolution of the EL mapping (see respective section) is taken into account, the current densities might be lower about a factor of eight.

Also the behavior of the integrated EL intensity as a function of the total current density strongly differs between the two samples, as shown in Fig. 4(c). The integrated EL intensity of the LED with ITO rises faster and more linearly than the one of the sample with Ni/Au top-contact. Dividing the integrated EL intensity by the corresponding current density, we also derived the relative external quantum efficiency (EQE), represented in Fig. 4(d). Again it is possible to identify several differences between the two samples. In particular, the relative EQE of the LED with ITO increases fast with current, and then almost saturates for current densities higher than 30 A/cm$^2$. In contrast, the relative EQE of the sample with Ni/Au contact increases slowly, and does not saturate in the studied range of currents. The latter trend is similar to the one reported by other groups for comparable devices contacted with Ni/Au/ITO;[12,16] in these cases, the relative EQEs saturate for much higher current densities (in the range 100–300 A/cm$^2$). This comparison, together with the other results previously discussed, suggests that the presence of a thin metal layer between the ITO and the p-GaN NW tips does not improve the overall characteristics of the contact. More importantly, our measurements also show that the LED with ITO exhibits a relative EQE about an order of magnitude higher than the sample with Ni/Au. This significant improvement may be explained by a better injection of holes together with a higher number of individual emitters and an enhanced extraction efficiency.

In conclusion, exclusively employing ITO instead of Ni/Au as the top contact for (In,Ga)N/GaN NW-LEDs on Si significantly improves many of the factors that have so far limited device performance; such as low number density of emitting NWs, poor hole injection efficiency, and high turn-on voltage. Furthermore, the low density of emitting NWs reported so far can be explained by a poor p-type top contact and is not caused by limitations intrinsic to (In,Ga)N/GaN NWs grown on Si.

The authors gratefully acknowledge D. Chatzopoulos for the analysis of the EL spots, B. Drescher as well as W. Anders for their help with the fabrication of the LEDs. This work has been partially supported by the European Commission (project DEEPEN, FP7-NMP-2013-SMALL-7, grant agreement no. 604416).

[1] A. Kikuchi, M. Kawai, M. Tada, and K. Kishino, Jpn. J. Appl. Phys. **43**, L1524 (2004).
[2] H.-M. Kim, Y.-H. Cho, H. Lee, S. I. Kim, S. R. Ryu, D. Y. Kim, T. W. Kang, and K. S. Chung, Nano Lett. **4**, 1059 (2004).
[3] S. Li and A. Waag, J. Appl. Phys. **111**, 071101 (2012).
[4] F. Glas, Phys. Rev. B **74**, 2 (2006).
[5] S. D. Hersee, A. K. Rishinaramangalam, M. N. Fairchild, L. Zhang, and P. Varangis, J. Mater. Res. **26**, 2293 (2011).
[6] C. Kölper, M. Sabathil, F. Römer, M. Mandl, M. Strassburg, and B. Witzigmann, Phys. Status Solidi A **209**, 2304 (2012).
[7] H. Sekiguchi, K. Kato, J. Tanaka, A. Kikuchi, and K. Kishino, Phys. Status Solidi A **205**, 1067 (2008).
[8] H.-W. Lin, Y.-J. Lu, H.-Y. Chen, H.-M. Lee, and S. Gwo, Appl. Phys. Lett. **97**, 073101 (2010).
[9] W. Guo, M. Zhang, A. Banerjee, and P. Bhattacharya, Nano Lett. **10**, 3355 (2010).
[10] A.-L. Bavencove, G. Tourbot, E. Pougeoise, J. Garcia, P. Gilet, F. Levy, B. André, G. Feuillet, B. Gayral, B. Daudin, and L. S. Dang, Phys. Status Solidi A **207**, 1425 (2010).
[11] R. Armitage and K. Tsubaki, Nanotechnology **21**, 195202 (2010).
[12] H. P. T. Nguyen, S. Zhang, K. Cui, X. Han, S. Fathololoumi, M. Couillard, G. a. Botton, and Z. Mi, Nano Lett. **11**, 1919 (2011).
[13] F. Limbach, C. Hauswald, J. Lähnemann, M. Wölz, O. Brandt, A. Trampert, M. Hanke, U. Jahn, R. Calarco, L. Geelhaar, and H. Riechert, Nanotechnology **23**, 465301 (2012).
[14] Y.-H. Ra, R. Navamathavan, J.-H. Park, and C.-R. Lee, ACS applied materials & interfaces **5**, 2111 (2013).
[15] A.-L. Bavencove, G. Tourbot, J. Garcia, Y. Désières, P. Gilet, F. Levy, B. André, B. Gayral, B. Daudin, and L. S. Dang, Nanotechnology **22**, 345705 (2011).
[16] W. Guo, A. Banerjee, P. Bhattacharya, and B. S. Ooi, Appl. Phys. Lett. **98**, 193102 (2011).
[17] H. P. T. Nguyen, S. Zhang, A. T. Connie, M. G. Kibria, Q. Wang, I. Shih, and Z. Mi, Nano Lett. **13**, 5437 (2013).
[18] E. Calleja, J. Ristić, S. Fernández-Garrido, L. Cerutti, M. A. Sánchez-García, J. Grandal, A. Trampert, U. Jahn, G. Sánchez, A. Griol, and B. Sánchez, Phys. Status Solidi B **244**, 2816 (2007).
[19] L. Geelhaar, C. Chèze, B. Jenichen, O. Brandt, C. Pfüller, S. Münch, R. Rothemund, S. Reitzenstein, A. Forchel, T. Kehagias, P. Komninou, G. P. Dimitrakopulos, T. Karakostas, L. Lari, P. R. Chalker, M. H. Gass, and H. Riechert, IEEE J. Sel. Topics Quantum Electron. **17**, 878 (2011).
[20] See supplemental material at [URL will be inserted by AIP] for further data and discussion of the comparability of the considered samples and of the leakage current in the NW-LEDs.
[21] A. M. Herrero, P. T. Blanchard, A. Sanders, M. D. Brubaker, N. Sanford, A. Roshko, and K. A. Bertness, Nanotechnology **23**, 365203 (2012).
[22] Y.-J. Lee, C.-J. Lee, C.-H. Chen, T.-C. Lu, and H.-C. Kuo, IEEE J. Sel. Topics Quantum Electron. **17**, 985 (2011).
[23] C.-H. Wang, Y.-W. Huang, S.-E. Wu, and C.-P. Liu, Appl. Phys. Lett. **103**, 233113 (2013).

[24] J. M. Shah, Y.-L. Li, T. Gessmann, and E. F. Schubert, J. Appl. Phys. **94**, 2627 (2003).